\documentclass[aps,prb,twocolumn,showpacs,superscriptaddress,floatfix]{revtex4-1}

\usepackage[utf8]{inputenc}
\usepackage{braket}
\usepackage{amsmath,amssymb,graphicx}
\usepackage{qcircuit}
\usepackage{color}

\usepackage[dvipsnames]{xcolor}

\begin{document}

\title{Quantum computation of magnon spectra}

\author{Akhil Francis}
\affiliation{Department of Physics, North Carolina State University, Raleigh, North Carolina 27695, USA}

\author{J.~K.~Freericks}
\affiliation{Department of Physics, Georgetown University, 37th and O Sts. NW, Washington, DC 20057 USA}

\author{A.~F.~Kemper}
\email{akemper@ncsu.edu}
\affiliation{Department of Physics, North Carolina State University, Raleigh, North Carolina 27695, USA}

\date{\today{}}

\begin{abstract}
We demonstrate quantum computation of two-point correlation functions for a Heisenberg spin chain.
Using  the IBM Q 20 quantum machines, we find that for two sites the correlation functions produce the exact results reliably.
For four sites, results from the IBM Q 20 Tokyo quantum computer are noisy due to read out errors and decoherence. Nevertheless, the correlation functions retain the correct spectral information. This is illustrated in the frequency domain by accurately extracting the magnon energies from peaks in the spectral function.  

\end{abstract}

\maketitle

\section{Introduction}
Interacting systems are typically characterized by properties of their ground state and of their low-lying excitations. For example, in spin systems the character of the low-energy excitations distinguishes a Heisenberg
model from an Ising or an XY model even when the ground states may be similar.
In quantum materials, the large variety of gapped
systems (that arise from charge-density waves, strong correlations, or superconductivity)
may be distinguished by carefully classifying their excitations.

The character of the low-energy excitations varies greatly depending on the physical behavior exhibited
by the material.  Consider an insulator whose low-energy behavior is described well by
interacting spins. It will exhibit different low-energy excitations than a metallic Fermi liquid, whose low-energy behavior is described well by electronic quasiparticles.  
Furthermore, different probes (such as optical conductivity, neutron scattering, or photoemission) probe different aspects of the system.
As a concrete example, consider
the low-energy excitations of the Fe-based superconductor FeSe. These have been viewed from both a
spin (neutron)\cite{wang_fese_neutron} and charge (optical)\cite{baum_fese} perspective. Both probes provide complementary information about the material. 

%

There are some many-body interacting systems that can have their spectrum analytically determined. 
In spin systems (like the XY model) a Holstein-Primakoff\cite{holstein1940field} or a Jordan-Wigner\cite{wigner1928paulische} transformation transforms the system into a form where the excitation spectrum is immediately determined. This occurs because the excitations of the spin system actually have a fermionic character that is cumbersome to extract in the original spin picture. Another approach is to guess the wave function and then obtain the excitations, e.g. as in BCS theory\cite{BCS} or in the quantum Hall effect.\cite{laughlin}
However, for a large class of systems no exact solution is known, and the correlation functions that encode the
low-energy excitations have to be obtained numerically. This may be achieved by a variety of
approaches including direct calculation through exact diagonalization (ED), many-body perturbation theory, density-matrix renormalization group, or quantum Monte Carlo (QMC) simulations. However, these either suffer from a finite size problem (as in ED or QMC),
or require other constraints (such as weak entanglement). An overview of these numerical methods is given in Ref.~\onlinecite{martin2016interacting}.

It is hoped that another potential solution to this problem can be found through the use of quantum computers. This will inevitably happen when reliable, fault-tolerant quantum computation is available on large size systems. We are not there yet, but quantum hardware is available now and in this work, we begin to illustrate how it can be employed for these types of problems.
Currently available quantum computers overwhelmingly work within a spin-qubit paradigm, where each qubit is a spin degree of freedom. Many-body problems involving spins are the most natural problems to consider on such hardware.

The current quantum computers, which have been termed noisy intermediate-scale quantum (NISQ) hardware\cite{preskill2018quantum}, are all constrained to have a small number of qubits that are of poor relative ``quality''; they have short coherence times and errors due to readout and imperfect gate application. This limits such hardware to small-scale problems with low depth circuits.  Much work has been devoted to both the improvement of the qubits and to their potential error correction, with the ultimate goal being quantum computers can implement fault-tolerant computation.
However, as we demonstrate here, 
accurate results can be obtained from the current generation of hardware,
when the calculations are performed robustly.

In this paper, we show how to calculate dynamical correlation functions using quantum computers. Our method is based on the work of Pedernales et al.\cite{pedernales2014efficient}, which we apply to two- and four-site Heisenberg spin models. We measure the spin-spin correlation functions by representing them in the time domain (after employing the Lehmann representation). This approach was also recently employed by Chiesa et al.\cite{chiesa2019quantum} within the context of molecular systems.

The remainder of the paper is organized as follows: In Sec.~II, We outline the calculational formalism, followed by quantum computations on the IBM 20-qubit Tokyo machine\cite{ibm} and IBM 20-qubit Almaden machine\cite{almaden}. The raw data are available online\cite{osfdata}.

\section{Formalism}
\begin{figure*}[htpb]
    \centering
      \includegraphics[width=0.9\textwidth]{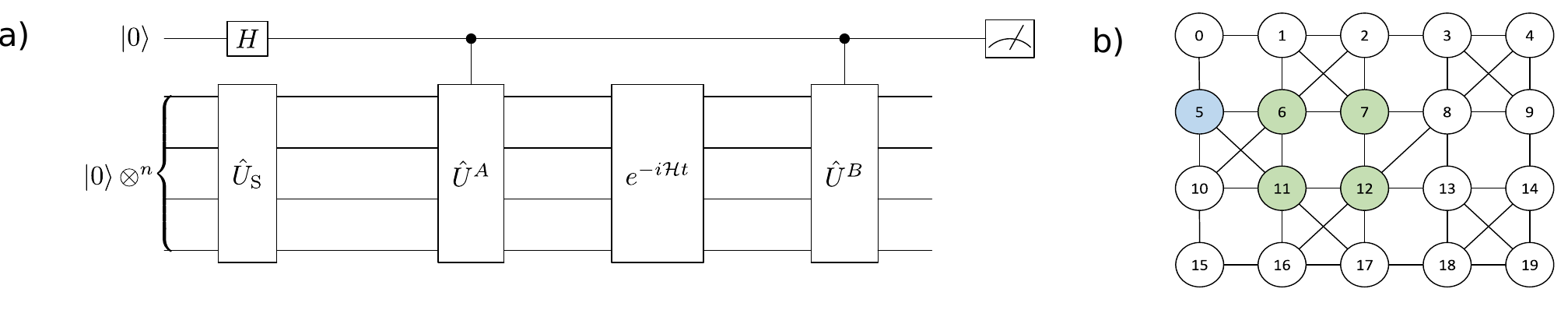}
   
    \caption{a) Quantum circuit for the two point correlator and b) layout used on the IBM Q 20 Tokyo machine. $H$ is the Hadamard rotation.
     }
    \label{fig:scheme}
\end{figure*}

In this work, we directly calculate the following two-point dynamical correlation function\cite{pedernales2014efficient}: 
\begin{align}
C(t)= \braket{\Phi|\hat{U}^B(t) \hat{U}^A(0)|\Phi},
\end{align}
between two unitary operators $\hat{U}^A$ and $\hat{U}^B$ evaluated at different times (in the Heisenberg representation). Note that these two operators are {\it not} the time evolution operators, but are typically different time-dependent spin operators.


The circuit we employ is shown in Fig.~\ref{fig:scheme}~(a) and follows the well-established strategy for evaluating such expectation values\cite{ortiz2001quantum}. First, the system is initialized in a particular state (such as a pure state that is a linear superposition $\ket{\Phi} = \sum_n c_n \ket{n}$ of energy eigenstates $\ket{n}$) via the application of a unitary operator $\hat U_{\rm s}$ onto the initial state of the quantum computer (which is the $|0\rangle^{\otimes n}$ state in the computational basis). In the work below, we choose $\ket{\Phi}$ to be the nondegenerate ground state in the antiferromagnetic case and the maximally polarized state for the ferromagnetic case. 
Then an ancilla qubit is employed to create an entangled state that entangles the two halves of the desired final matrix element of the system (bra and ket, here denoted $\bra{\Phi}$ and $\ket{\Phi}$) with the different states of the ancilla qubit (for an expectation value these system states are identical, while for a general matrix element, they can be different; here we compute an expectation value).
At this stage (after a Hadamard operation on the ancilla), the (pure) state stored in the quantum computer is
\begin{align}
\ket{\psi}=\frac{1}{\sqrt{2}}[\ket{0}\otimes  \ket{\Phi} + \ket{1}\otimes  \ket{\Phi}].
\end{align}
The second step is to apply a controlled-$U^A$ operation with the control on the ancilla qubit and the $U^A$ operating only on the system qubits
\begin{subequations}
\begin{align}
\ket{\psi}=&\frac{1}{\sqrt{2}} \ket{0}\otimes  \ket{\Phi} + \frac{1}{\sqrt{2}}  \ket{1} \otimes U^A \ket{\Phi} \\
=& \frac{1}{\sqrt{2}} \ket{0}\otimes  \ket{\Phi} + \frac{1}{\sqrt{2}} \ket{1} \otimes \sum_{mn} c_n \bra{m}U^A \ket{n} \ket{m} .
\end{align}
\end{subequations}
Here we have expanded 
\begin{align}
U^A \ket{n} = \sum_m \bra{m}U^A \ket{n} \ket{m},
\end{align}
with $\{\ket{m}\}$ and $\{\ket{n}\}$ both being complete sets of states for the {\it system} (the state $\ket{\Phi}$ is equal to $\sum_nc_n\ket{n}$).
The system is then evolved forward in time according to the Hamiltonian (via the operation $\exp(-i{\mathcal H}t/\hbar)$, followed by the controlled-$U^B$ operation, which yields
\begin{subequations}
\begin{align}
\ket{\psi}=&\frac{1}{\sqrt{2}}  \ket{0}\otimes \sum_m c_m e^{-iE_m t} \ket{m} \nonumber\\
+& \frac{1}{\sqrt{2}} \ket{1} \otimes \sum_{mn} c_n e^{-i E_m t} \bra{m}U^A \ket{n} U^B\ket{m} \\
=&\frac{1}{\sqrt{2}} \ket{0}\otimes \sum_m c_m e^{-iE_m t} \ket{m} \nonumber\\
+& \frac{1}{\sqrt{2}} \ket{1} \otimes
\sum_{lmn} c_n e^{-i E_m t}\bra{l}U^B \ket{m}  \bra{m}U^A \ket{n}\ket{l}.
\end{align}
\end{subequations}
Finally, measuring the ancilla qubit then determines the real and imaginary part of the correlation function. 

In particular, we first extract the reduced density matrix $\rho^{An}$ of the ancilla (by tracing out the system). The diagonal elements are equal to $\frac{1}{2}$, and the off-diagonal term is
\begin{align}
\rho^{An}_{0,1}&= \frac{1}{2} \sum_{lmn} c_l^*c_n e^{-i (E_m -E_l) t} \bra{l}U^B \ket{m}\bra{m}U^A \ket{n}. 
\label{eq:densmat}
\end{align}
This is precisely the Lehmann representation of the correlation function.
%
%
 %
%
%
 Hence, measuring the ancilla qubit in the $x$- [by applying a Hadamard gate] and $y$- [by applying a $R_x(-\pi /2)$ gate] bases, finally yields the real and imaginary parts of the desired correlation function. The projective measurement of the reduced density matrix in the x or y basis is given by 
\begin{align}
\Pr(\rho_{An})_{(\ket{0}\bra{0})}^{x,y}  = \frac{1}{2}\left[1+(\mathrm{Re,Im}) C(t)\right].
\label{eq:prefactors}
\end{align}
Below, we will be using Pauli matrices as the operators $U^{A/B}$, which restricts the real and imaginary parts of $\vert C(t) \vert \leq 1$; this
is compatible with the interpretation of the ancilla qubit measurement as a probability.

  


\section{Results}
Spin systems are a natural choice to study in digital quantum computers because they are
directly mapped onto qubits.
Work on spin systems has already begun by others\cite{chiesa2019quantum,smith2019simulating}. We continue this work on the
periodic Heisenberg model
\begin{align}
\mathcal{H} = J \sum_i \mathbf{S}_i \cdot \mathbf{S}_{i+1},
\end{align}
which is a representative model for a variety of magnetic systems. Here $J$ is the Heisenberg exchange integral and $\mathbf{S}_i$ is an SU(2) quantum spin operator at lattice site $i$ with components in the $x$, $y$, and $z$ directions.
Due to limitations of the current quantum hardware, applications are restricted to two-
and four-site models with periodic boundary conditions.
For these models, the low-energy excitations obtained from the correlation function are the transverse or longitudinal spin-spin correlation functions
\begin{align}
\langle S^i_\alpha(t) S^j_\alpha(0)\rangle,
\end{align}
where $\alpha \in \{x,y,z\}$ is the spatial component of the Pauli spin and $i,j$ indicate lattice sites. We will work in units where the
electron spin $\hbar/2$ is set to unity, so the spin operators ($ {S}_\alpha $) are the Pauli matrices.
Due to spatial translation invariance,
the correlation functions only depend on the distance between sites $\Delta r \equiv |i-j|$.

\subsection{Two-site Heisenberg model}
To demonstrate the feasibility of our approach, we first
apply our methodology to the two-site anti-ferromagnetic Heisenberg model with the unitary operators $U^{A/B} = S_z$. The two site Hamiltonian, $\mathcal{H}_{12}$ is defined as
\begin{align}
\mathcal{H}_{12} = J  \mathbf{S}_1 \cdot \mathbf{S}_2,
\end{align}
The time evolution operator $\exp\left(-i \mathcal{H}_{12} t\right)$ is implemented as shown in Fig.~\ref{fig:circuit_2site}, which is based on a Cartan
KAK decomposition\cite{vidal2004universal}. For this
\begin{figure}[h]
\vspace{0.1in}
\scalebox{.7}{
\Qcircuit @C=1em @R=.7em {
&\qw & \ctrl{1}	& \qw  & \gate{R_x(-2Jt-\pi/2)}  & \qw  & \gate{H}& \ctrl{1} & \qw&\gate{H} &\qw&\qw &\ctrl{1} &\qw&\gate{R_x(\pi/2)} & \qw  \\
&  \qw  &\gate{X}& \qw  & \gate{R_z(-2Jt)} &\qw &\qw &\gate{X} &\qw & \gate{R_z(2Jt)} &\qw &\qw &\gate{X}&\qw&\gate{R_x(-\pi/2)}&\qw 
}
}
\caption{Quantum circuit for the evaluation of $\exp\left(-i \mathcal{H}_{12} t\right)$. $X$ denotes a Pauli X, while $R_\alpha(\theta)$ denotes a rotation by $\theta$ about the axis $\alpha$. }
\label{fig:circuit_2site}
\end{figure}
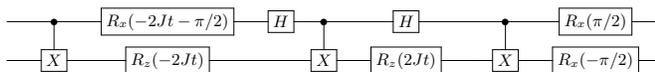
\begin{figure}[htpb]
    \centering
     \includegraphics[width=0.49\textwidth]{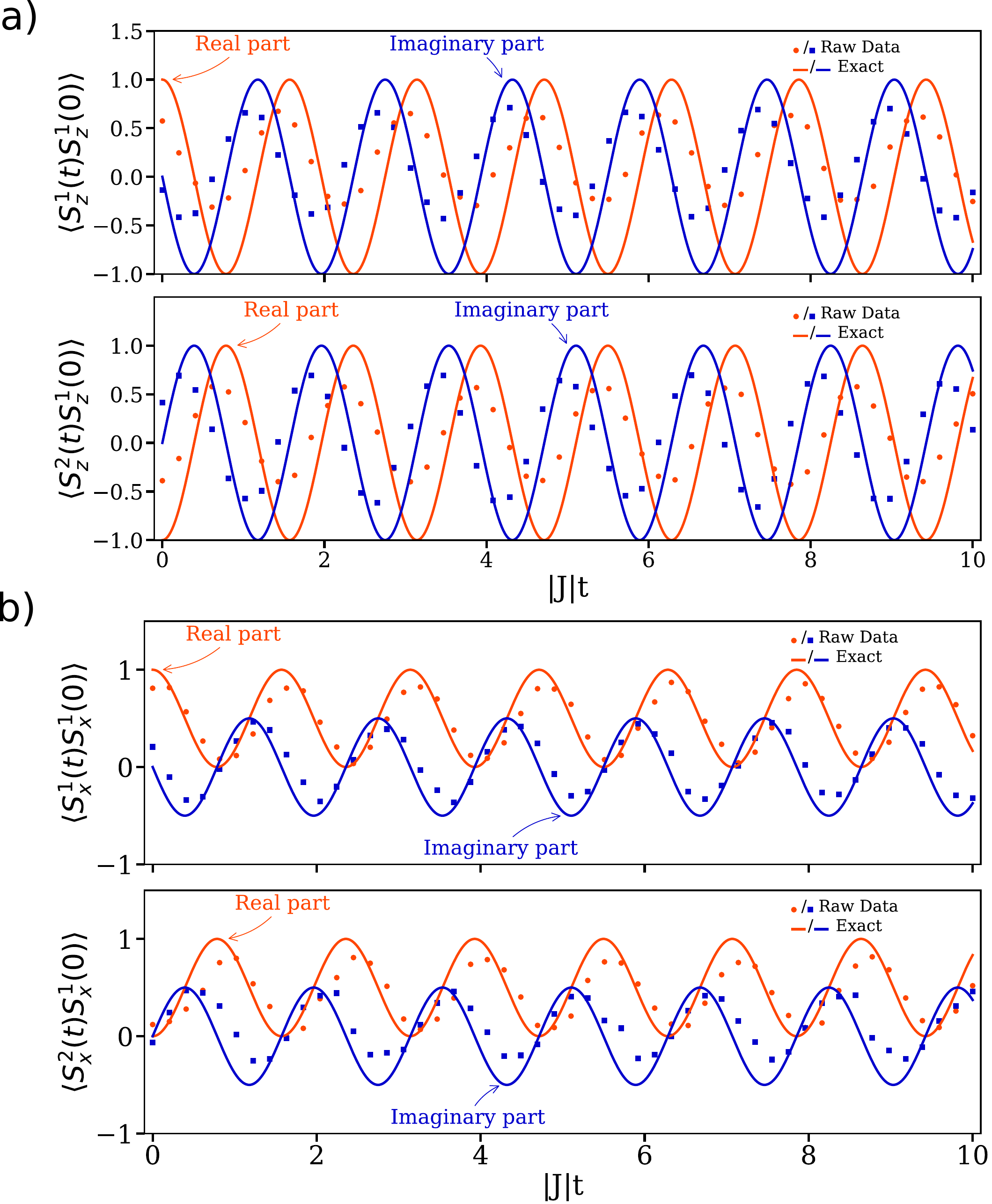}
    \caption{Real and imaginary parts of the (a) $zz$ (b) $xx$ correlation function for the two-site (a) antiferromagnetic and (b) ferromagnetic Heisenberg model. Solid lines are the exact solution, circles and squares are the results obtained from the IBM Q hardware.  Note that the period of the quantum computation is well reproduced, even though the amplitude of the oscillations is reduced.
    The computations were performed on IBM Q Tokyo and Almaden for (a) and (b), respectively.
    }
    \label{fig:2site_ferro_r0}
\end{figure}
particular system, the time evolution can be executed without Trotterization and 
the time simply enters as a parameter in the rotation gates. In other words, one can implement the time evolution for any time with the same number of gate operations. This occurs only because of special features of the model and this small cluster; it will not scale.
Fig.~\ref{fig:2site_ferro_r0} a)
shows the results for the $zz$ spin-spin correlation function $\langle S^i_z(t)S^j_z(0)\rangle$ compared to the exact solution. Although the results from the IBM Q 20 Tokyo machine have an amplitude that is closer to random noise than the expected value, the measurements show a faithful reproduction of the period of the oscillations found in the analytic results.  In a similar way ferromagnetic $xx$ spin-spin correlation function has been calculated and the results obtained from IBM Q Almaden is shown in Fig.~\ref{fig:2site_ferro_r0} b). 

\subsection{Four-site Heisenberg model}

%

\begin{figure}[h]
\includegraphics[width=0.49\textwidth]{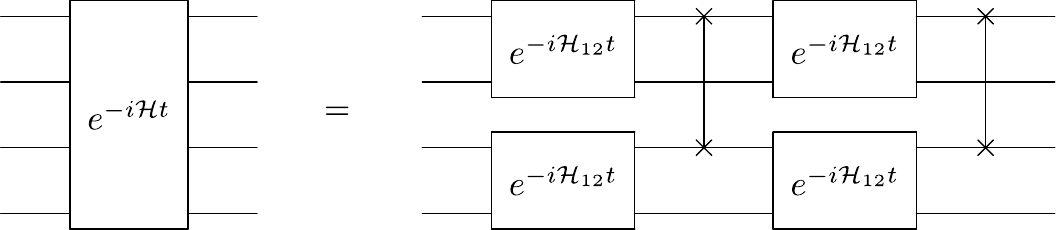}
\caption{Quantum circuit for the evaluation of $\exp\left(-i \mathcal{H} t\right)$ for a 4-site ferromagnetic Heisenberg chain. For this particular system,
the full time evolution can be factorized into four pairwise applications of $\exp\left(-i \mathcal{H}_{12} t\right)$.
Due to the topology of the layout, swap gates may be eliminated.}
\label{fig:circuit_4site}
\end{figure}

%
Next, we extend the circuit to a four-site model, where we compute the $xx$ spin-spin
correlation function $\langle S^i_x(t)S^j_x(0)\rangle$ for the ferromagnetic ground state.
The ferromagnetic ground state is a computational basis product state, which minimizes the number of gates required for the calculation.
Here we have broken the $SU(2)$ symmetry of the Hamiltonian by choosing the ground state to be one (all aligned along the $z$ direction) of the ferromagnetic multiplet of ground states. 
In this case, we can also carry out the time evolution without Trotterization;
we time evolved the state in a pairwise manner between sites (see Fig.~\ref{fig:circuit_4site}).
The time evolution is factored as
\begin{align}
\exp\left(-i J t \sum_i \mathbf{S}_i\cdot \mathbf{S}_{i+1}\right) &= \nonumber \\
\exp\left(-iJt \mathbf{S_1} \cdot \mathbf{S_2}\right)& \exp\left(-iJt \mathbf{S_3} \cdot \mathbf{S_4}\right) \times \nonumber \\
\exp\left(-iJt \mathbf{S_4} \cdot \mathbf{S_1}\right) &\exp\left(-iJt \mathbf{S_2} \cdot \mathbf{S_3}\right).
\end{align}
Although in general this way of time evolving is not accurate due to non-commutativity of operators, for this particular situation it turns out to be exact due to a fortuitous cancellation of terms. For four sites this decomposition is exact in the all spin aligned and one spin flipped sub sector. In the case of ferromagentic ground state xx correlation function, this is the sub sector that is relevant. For other ground states or larger systems one needs to incorporate the full Trotterization scheme for the correct time 
evolution.

\begin{figure}[htpb]
    \centering
    \includegraphics[width=0.49\textwidth, clip=true, trim=0 30 40 80]{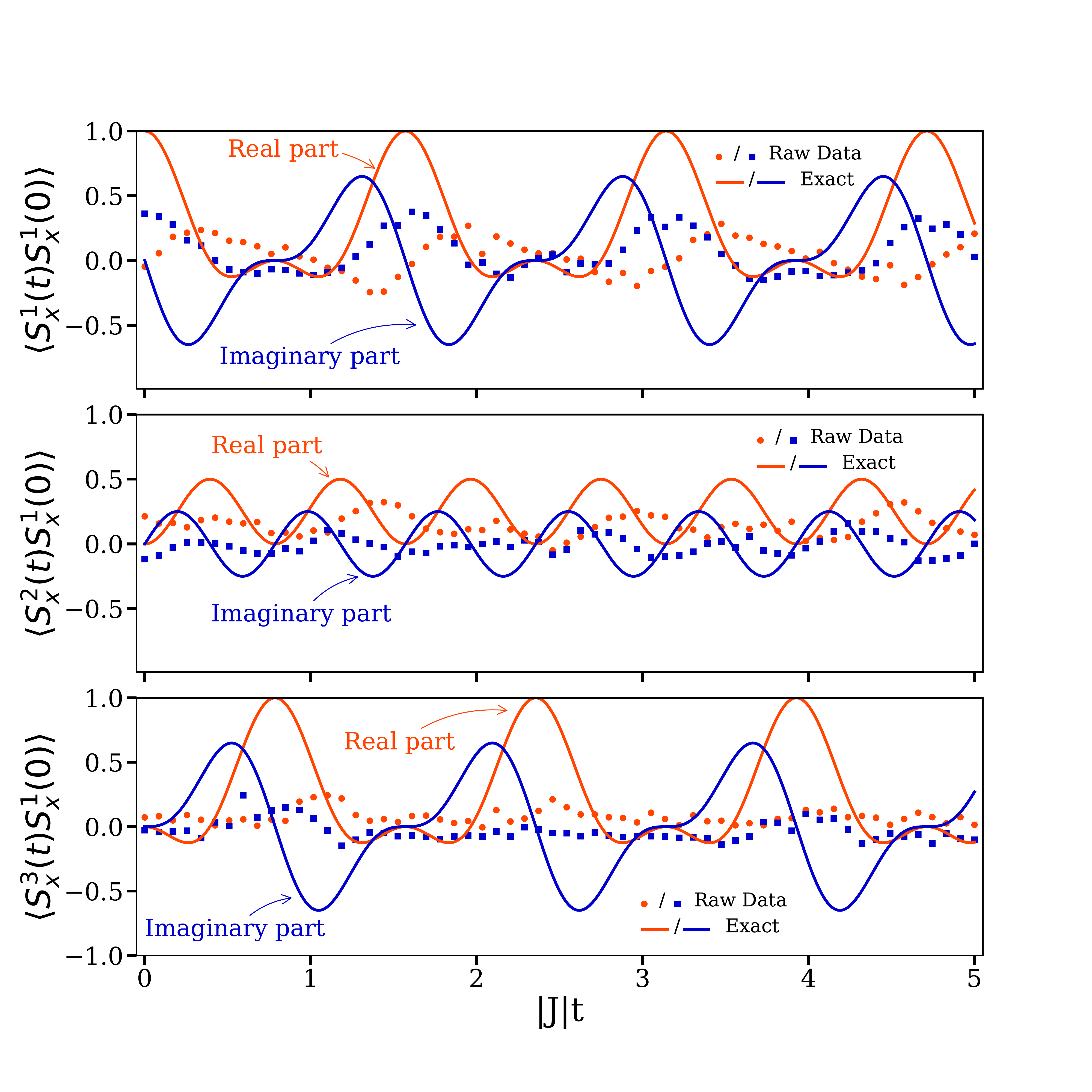}

    \caption{$xx$ spin-spin correlation function for the four-site ferromagnetic Heisenberg model.
    The measured raw data (circles, from the IBM Q 20 Tokyo) are compared with the analytic (solid lines) results. 
     }
    \label{fig:4site_actual_data}
\end{figure}
Fig.~\ref{fig:4site_actual_data} shows the measured results from
the IBM Q 20 Tokyo for the three possible values of $\Delta r$ on a four-site cluster, as well as the analytic result for benchmarking. We have exploited the geometry of the IBM Q 20 machine to avoid as many swap gates as possible; the four sites were laid out in a circular pattern, and the ancilla was directly connected
to two of the sites (see Fig.~\ref{fig:scheme}).
In stark contrast to the two-site model, the measured data are far from the exact results.  However, some patterns may still be recognized. To improve the quality of the data, we employ two types of error mitigation techniques.

First a readout error mitigation is applied to the data, where the readout error of the ancilla qubit is calibrated based on the measurement of a pure $\ket{0}$ or $\ket{1}$ state. If the ancilla qubit is measured without any gates applied, ideally the probability to obtain state $ \ket{1} $, $p(1|0)$ is 0 and  probability to obtain $ \ket{0}$ state, $p(0|0)$ is 1, and similar for the measurement after applying an $X$ gate.
In practicality, these probabilities will have some noise. A calibration matrix can be defined as 

\begin{equation}
C =
\left(\begin{array}{cc} p(0|0) & p(0|1)  \\ p(1|0) & p(1|1) \end{array}\right)
\end{equation}

Thus in order to get the read out corrected value of the ancilla qubit measurements we need to multiply by the inverse of the caliberation matrix.
\begin{equation}
\left(\begin{array}{c} p_{0,\mathrm{corrected}} \\ p_{1,\mathrm{corrected}} \end{array}\right) =
(C) ^{-1}
\left(\begin{array}{c} p_{0,\mathrm{measured}} \\ p_{1,\mathrm{measured}} \end{array}\right) 
\end{equation}

Second, a {\it phase and scale} correction is also applied. This technique is already known to be applied in the context of spin dynamics\cite{chiesa2019quantum}. Using the exact result for the equal-time onsite spin-spin correlation function (in our units)
\begin{align}
\langle S^1_x(0) S^1_x(0)\rangle = 1,
\end{align}
we determine a complex multiplicative correction factor by comparing the measured complex correlation function to this exact result.
All the data points are then multiplied by this complex factor to correct their phase and scale.
We do not know exactly why the same phase and scale factor improves all the data points, it could be correcting the asymmetry in measurement error or correcting the errors coming from the unitary gates which has the same structure for all the data points. This correction seems to work empirically.

\begin{figure}[h]
    \centering
    
   
\includegraphics[width=0.49\textwidth,clip=true, trim=0 0 0 0]{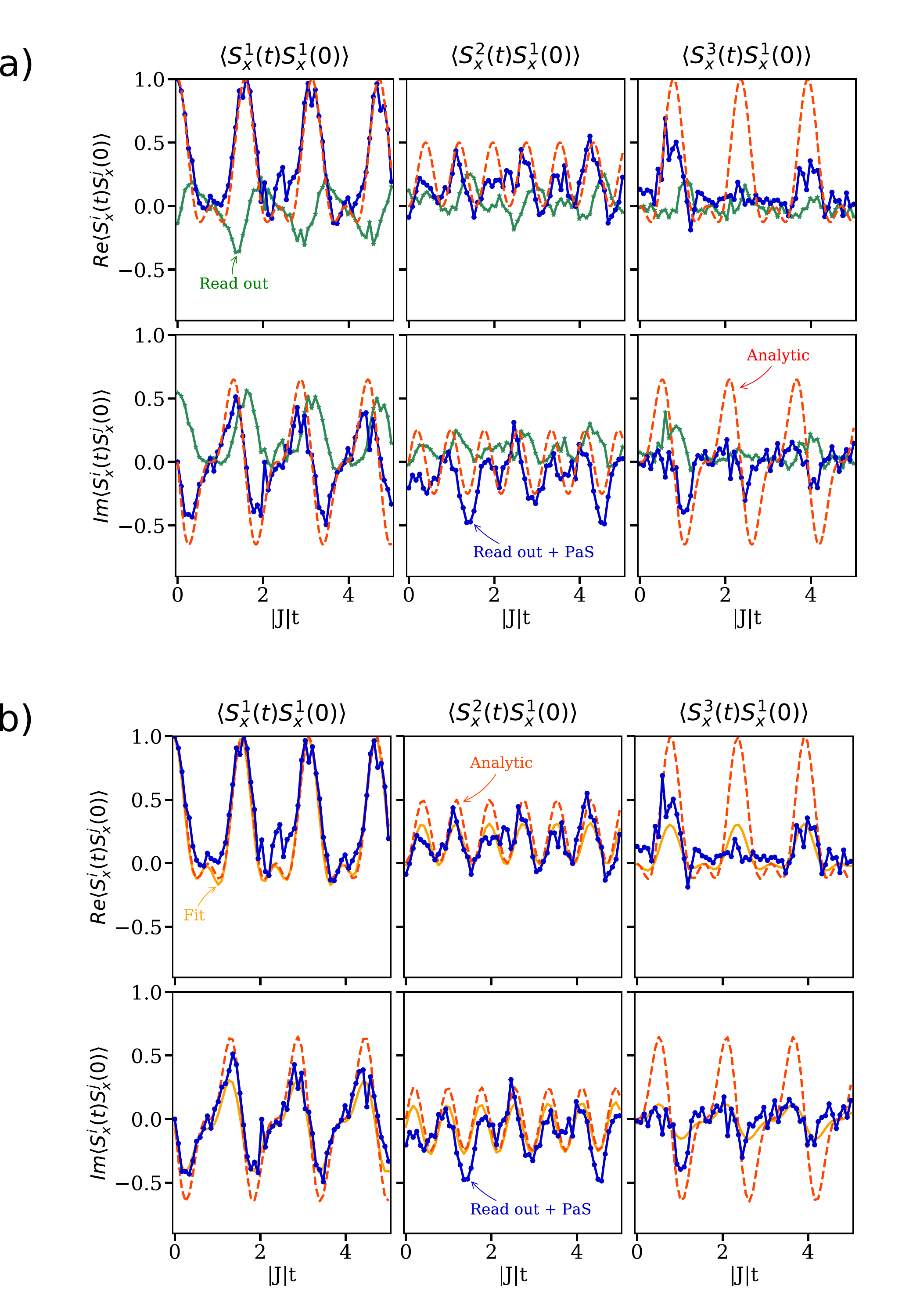}
    \caption{$xx$ spin-spin correlation function. a) Comparison of the error-mitigated data after just a readout correction (green), with both a read out and phase-and-scale (labelled PaS) correction (blue) and the corresponding analytical result (red). 
    b) Comparison of the phase-and-scale and readout corrected data (blue), the Fourier fit as discussed in the main text (orange) and the  analytic result (red).  The top and bottom rows show the real and imaginary parts of the function respectively. Clearly, full error mitigation is required to extract meaningful results.}
    \label{fig:PaS_data}
\end{figure}

The error-mitigated data is shown in Fig.~\ref{fig:PaS_data}. While the readout error mitigation makes some minor changes, a clear effect can be seen due to the phase-and-scale mitigation.  In particular, the $\Delta r=0$ data now shows a signal that is clearly reflective of the analytic results; the
remaining $\Delta r\neq 0$ also exhibit some oscillations, although not as clearly.


\begin{figure}[htb]
     \includegraphics[width=0.4\textwidth, clip=true, trim= 0 80 20 160]{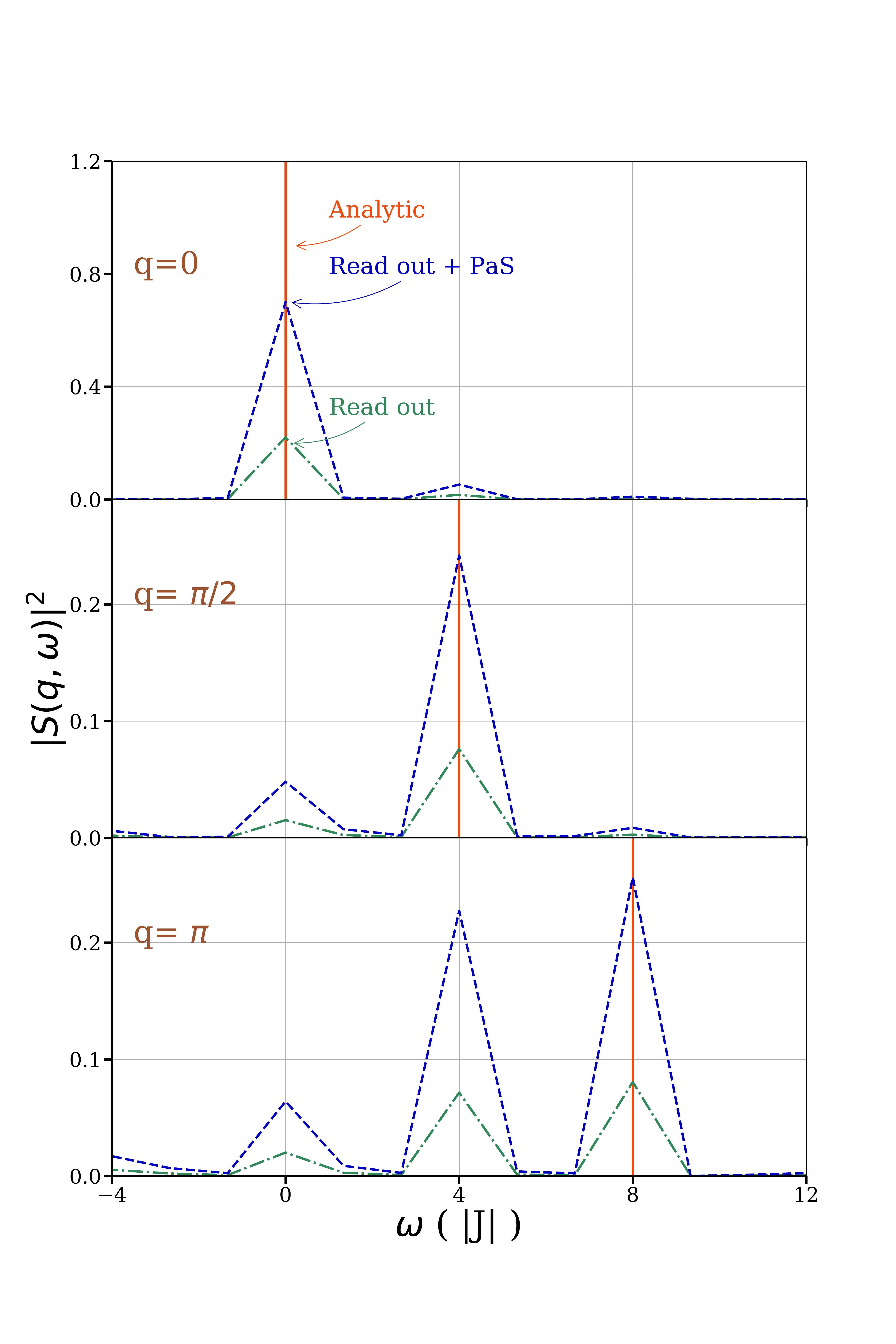}
    
    \caption{The dynamic spin susceptibility, $|S(q,\omega)|^2$. It is obtained from the $xx$ spin-spin correlation functions for the four-site ferromagnetic Heisenberg model. Readout corrected, and readout+phase-and-scale corrected data are compared with the analytic result. 
    }
    \label{fig:4site_Sqw_abs}
\end{figure}

The low-energy excitations of the periodic Heisenberg chain are magnons.
A reconstruction of the magnon spectrum requires a Fourier transform of the correlation function
from real space to momentum space, and from time to frequency. This is because the magnon dispersion is extracted as the peak of the corresponding dynamical spin susceptibility. Although
the Fourier transformations may be performed directly, an issue arises due to the 
inequivalent noise between the different measurements. The Fourier
transform relies on an interference between different correlation functions, and if those terms
have different amplitudes due to variable noise, contamination across channels may occur.
We demonstrate this below; in preparation for that discussion,
we first consider a different treatment of the data. Based on the assumption
of a single frequency $\omega_q$ for each $q$ point, we globally fit the measured
data for all $\Delta r$ with an inverse Fourier transform of this assumption, letting
the amplitudes $A_q$ and frequencies $\omega_q$ be free variables:
\begin{align}
\langle S^i_x(t) S^j_x(0)\rangle = \sum_q A_q e^{i q\left(r_i - r_j \right)} e^{-i \omega_q t}
\end{align}
In the absence of noise, the $A_q$ are given by the usual constants for a $4$-site Fourier transform;
with the noise present in the quantum computer, these are not exact.
The results are shown in Fig.~\ref{fig:PaS_data}
together with the phase-and-scale mitigated data and the analytic solution.




\begin{figure*}[htb]
    \centering

    \includegraphics[width=0.4\textwidth,clip=true, trim=30 330 0 0]{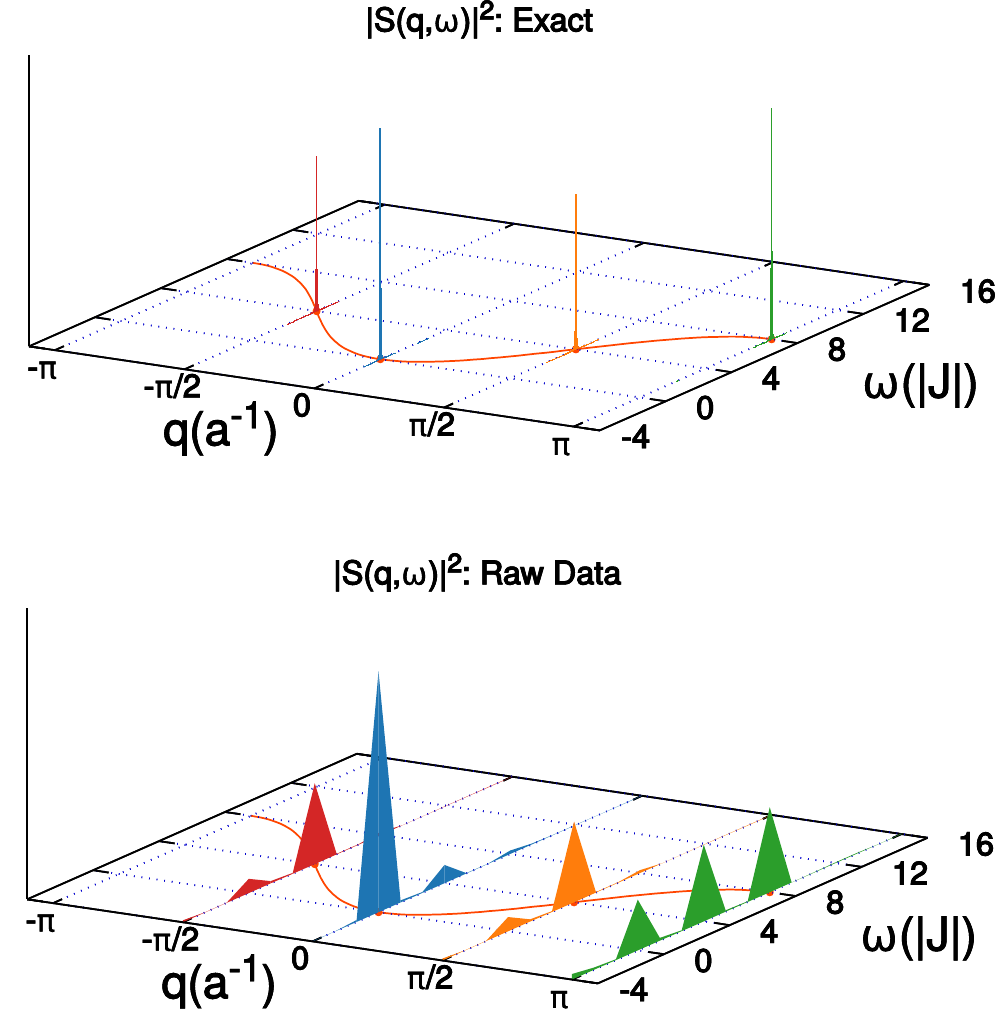}
    \includegraphics[width=0.4\textwidth,clip=true, trim=30 0 0 320]{3Drev.pdf}

    \caption{$\vert S(q,\omega)\vert^2$ for all the four cases along with the magnon dispersion curve. The fit plot has peaks at the frequencies obtained by fitting the data. The raw-data and phase-and-scale error mitigated plots have a finite width because of the finite number of data points in the time domain. The raw-data plot is scaled up. In all cases, peaks at the expected magnon frequencies are clearly visible;  for the raw-data and phase-and-scale corrected data there is leakage to other channels.}
    
    \label{fig:3D}
\end{figure*}

Fig.~\ref{fig:4site_Sqw_abs} plots the power spectra of the spin-spin correlation functions in energy and momentum $|S(q,\omega)|^2$, including the analytic
results, the raw data and the readout+phase-and-scale mitigated data. The expected results,
which are unique peaks in the spectra for the magnon excitations, are clearly seen in the spectrum of
the analytic results. The raw data from the quantum calculation, while they
appear extremely noisy, do contain the spectral content of the magnons; We
observe peaks at the correct frequencies; however, there is some contamination
between the channels. All momenta have signal at $\omega=0$, and there is
notable content from $q=0$ at finite frequency.  

As mentioned above, this contamination may be ameliorated by a Fourier transform with the
amplitudes and frequencies as fitting parameters.
The results
are shown in Fig.~\ref{fig:PaS_data}. The obtained fitted frequencies are -0.03,  4.03,  7.97,  4.03 while the exact values are 0,4,8,4 respectively.

    

    


\section{Conclusion}
We have outlined and applied a quantum circuit for evaluating the low-energy excitations of the
periodic Heisenberg chain.
For a very short chain (two-site), one obtains reasonable behavior of the system; for an intermediate
length chain (four-site), the raw data contains significant noise. Nevertheless, we can still obtain the correct frequency information from the data. This is further improved by applying a phase-and-scale correction for each set of measurements of the correlation
function.
Our results suggest that this approach of computing correlation functions for space/time 
translation-invariant systems, or more generally properties that can
be expressed as interference patterns for systems with these kinds of symmetries,
may not require fault-tolerant computation.  Rather, the Fourier transforms act as effective filters that naturally
enhance the oscillation patterns observed in the data.  In this work, we have shown that this is the case for
small systems, but with the availability of higher quality qubits this approach can be expanded to even larger systems.

\begin{acknowledgments}
This work was supported by the Department of Energy, Office of Basic Energy Sciences, Division of Materials Sciences and Engineering under Grant No. DE-SC0019469. J. K. F. was also supported by the McDevitt bequest at Georgetown.We acknowledge the use of IBM Q via the IBM Q Hub at NC State for this work. The views expressed are those of the authors and do not reflect the official policy or position of the IBM Q Hub at NC State, IBM or the IBM Q team. We acknowledge the use of Qiskit software package\cite{Qiskit} for doing the quantum simulations.
\end{acknowledgments}



\bibliography{ref}

\appendix*
\section{Four site time evolution }
For the four site time evolution, we have decomposed the circuit into pairwise qubit time evolution as
\begin{align}
    \exp(-iHt) =& \exp(-iH_{12}t)\exp(-iH_{34}t)\nonumber\\
    & \times \exp(-iH_{23}t)\exp(-iH_{14}t)
\end{align}{}
where $H = H_{12} + H_{23} + H_{34} + H_{14} $ and $H_{ij} $ is the two site Heisenberg interaction Hamiltonian. This decomposition is not exact in general, but for all spin aligned up(m=4) and three spins aligned up and one spin along down (m=2) sectors this decomposition is exact. For calculating two point correlation functions in ferromagnetic ground state only these sectors matter. This is the reason why we do not need to use Trotterisation in our four site calculations. The resulting time evolution matrix for m=2 sector is
\begin{align}
U_{m=2}(t) &=
\left(
\begin{array}{cccc} 
a_0 & a_1 & a_2 & a_1  \\ 
a_1 & a_0 & a_1 & a_2 \\
a_2 & a_1 & a_0 & a_1 \\
a_1 & a_2 & a_1 & a_0
\end{array}
\right),
\end{align}
where
\begin{align}
a_0 = \cos(2Jt)^2,\nonumber \hspace{1mm}
a_1 = -\frac{i}{2} \sin(4Jt),\nonumber \hspace{1mm}
a_2 = -\sin(2Jt)^2.\nonumber
\end{align}


\end{document}